    \documentclass[lettersize,journal]{IEEEtran}
    \usepackage{booktabs}
    \usepackage{diagbox}
    \usepackage{makecell}
    \usepackage{array}
    \usepackage{graphicx,amssymb,amsmath}
    \usepackage{multicol}
    \usepackage[noadjust]{cite} 
    \usepackage{setspace}
    \usepackage{subfigure}
    \usepackage{graphicx}
    \usepackage{float}
    \usepackage {url}
    \usepackage{stfloats}
    \usepackage{amsthm,pifont}
    \usepackage{flushend}
    \usepackage{cases,subeqnarray}
    \usepackage{bm,multirow,bigstrut}
    \usepackage{amsmath, amsthm, amssymb}
    \usepackage{textcomp}
    \usepackage{latexsym,bm}
    \usepackage{booktabs}
    \usepackage{xcolor}
    \usepackage{mathtools}
    \usepackage{dsfont}
    \usepackage{extarrows}
    \usepackage{epsfig}
    \usepackage{epsfig}
    \usepackage{epstopdf}
    \PassOptionsToPackage{options}{xcolor}
    \usepackage{colortbl}
    \usepackage{multirow, hhline}
    \usepackage{array} 
    \usepackage{algorithmicx,algorithm}
    
    \theoremstyle{plain}

    \theoremstyle{plain}

    \usepackage{amsmath}

    \IEEEoverridecommandlockouts
    \begin{document}
    \title{Generative AI for Integrated Sensing and Communication: Insights from the Physical Layer Perspective}
    
    \author{
    Jiacheng Wang,
    Hongyang Du,
    Dusit Niyato,~\IEEEmembership{Fellow,~IEEE}, Jiawen Kang, Shuguang Cui,~\IEEEmembership{Fellow,~IEEE}, \\ Xuemin (Sherman) Shen,~\IEEEmembership{Fellow,~IEEE}, and Ping Zhang,~\IEEEmembership{Fellow,~IEEE} 
    \thanks{J.~Wang, H.~Du and D. Niyato are with the School of Computer Science and Engineering, Nanyang Technological University, Singapore (e-mail: jiacheng.wang@ntu.edu.sg, hongyang001@e.ntu.edu.sg, dniyato@ntu.edu.sg).}
    \thanks{J. Kang is with the School of Automation, Guangdong University of Technology, Guangzhou, China (e-mail: kavinkang@gdut.edu.cn).}
    \thanks{S. Cui is with the School of Science and Engineering (SSE) and the Future Network of Intelligence Institute (FNii), Chinese University of Hong Kong (Shenzhen), China (e-mail: shuguangcui@cuhk.edu.cn).}
    \thanks{X. Shen is with the Department of Electrical and Computer Engineering, University of Waterloo, Canada (email: sshen@uwaterloo.ca).}
    \thanks{P. Zhang is with the State Key Laboratory of Networking and Switching Technology, Beijing University of Posts and Telecommunications, China (e-mail: pzhang@bupt.edu.cn).}
    }
\maketitle
    \begin{abstract}
    As generative artificial intelligence (GAI) models continue to evolve, their generative capabilities are increasingly enhanced and being used extensively in content generation. Beyond this, GAI also excels in data modeling and analysis, benefitting wireless communication systems. In this article, we investigate applications of GAI in the physical layer and analyze its support for integrated sensing and communications (ISAC) systems. Specifically, we first provide an overview of GAI and ISAC, touching on GAI's potential support across multiple layers of ISAC. We then concentrate on the physical layer, investigating GAI's applications from various perspectives thoroughly, such as channel estimation, and demonstrate the value of these GAI-enhanced physical layer technologies for ISAC systems. In the case study, the proposed diffusion model-based method effectively estimates the signal direction of arrival under the near-field condition based on the uniform linear array, when antenna spacing surpassing half the wavelength. With a mean square error of 1.03 degrees, it confirms GAI's support for the physical layer in near-field sensing and communications.
    \end{abstract}
    \IEEEpeerreviewmaketitle
    \section{INTRODUCTION}
    Recently, the unprecedented growth in user data, together with the continuous advancement of AI-Generated Content (AIGC) models, have led to groundbreaking applications such as Google Bard and ChatGPT. As users increasingly benefit from these applications, their attention is concurrently shifting to the mechanism powering these applications, i.e., generative artificial intelligence (GAI)~\cite{bond2021deep}. Unlike traditional AI models that prioritize sample analysis, training, and classification, GAI specializes in understanding and modeling the distribution of complex datasets. By leveraging statistical properties of the training data, GAI can generate data similar to the training data, manifesting in diverse formats like documents and images~\cite{du2023beyond}. For example, the diffusion model-based ControlNet~\cite{zhang2023adding} can efficiently generate images with outstanding quality, in terms of resolution, saturation, and naturalness, according to the user prompts, which demonstrates greater flexibility and higher efficiency compared to traditional content generation methods. In the context of the rapidly evolving of wireless network services, GAI is poised to meet the various and ever-changing demands of users.

   Indeed, not only content generation, GAI's inference capability has catalyzed research across various domains. For example, researchers in~\cite{sun2020generative} introduce a generative adversarial networks (GANs) based architecture to learn the channel transition probabilities (CTP) from the observations, thereby helping achieve the maximum likelihood sequence detection. Additionally, in device-to-device (D2D) communications, an incentive mechanism based on contract theory is proposed to facilitate user information sharing, in which the diffusion model is employed to generate optimal contract designs~\cite{du2023beyond}. 
   
\begin{figure*}[htp]
    \centering
    \includegraphics[width=1\linewidth]{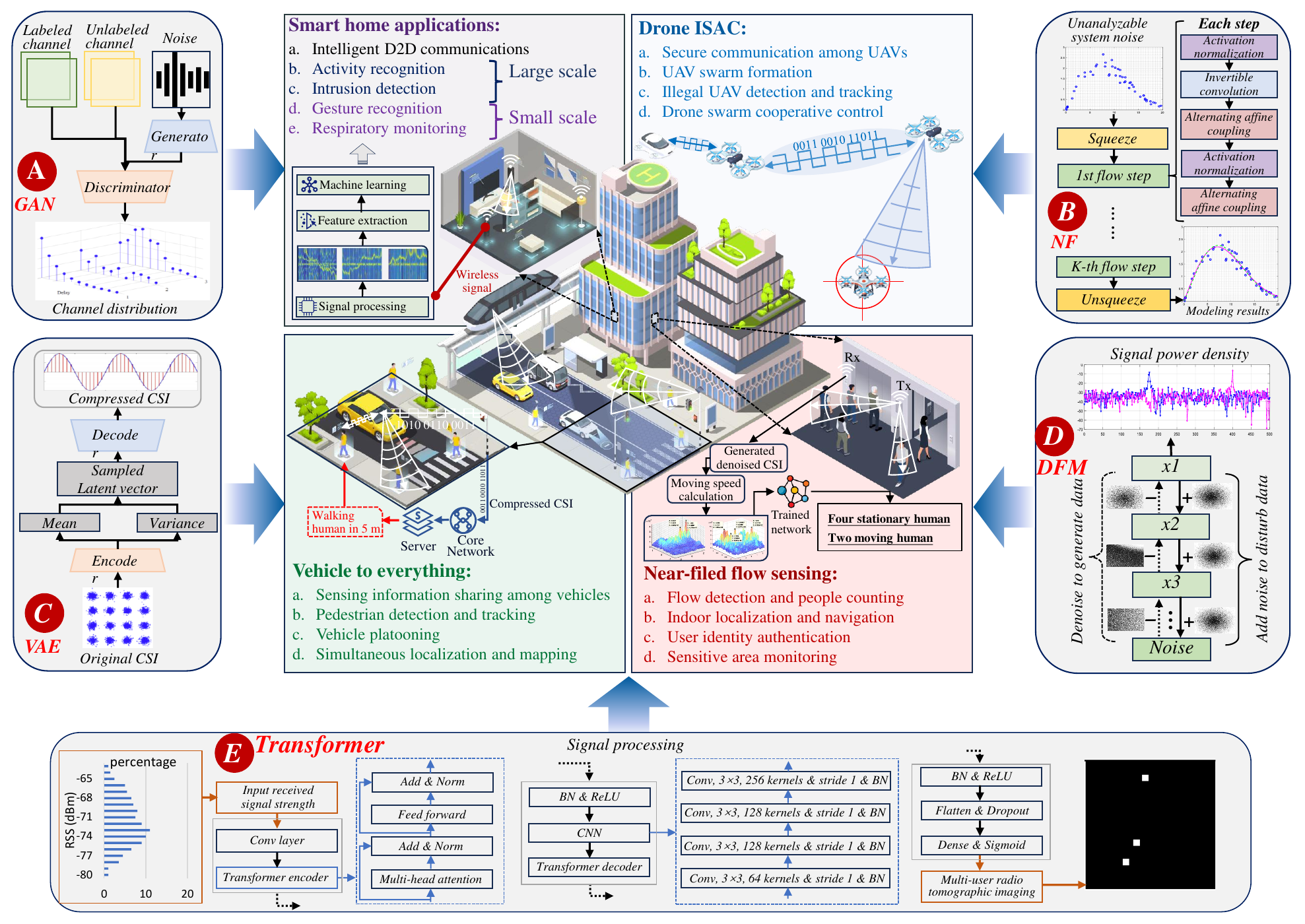}
    \caption{The role of GAI in the physical layer and its support for ISAC applications. The GAI models can be utilized to enhance several physical layer technologies, including channel state information (CSI) compression and signal detection. On this basis, the GAI enhanced physical layer technologies can further augment ISAC system performance across various applications, such as indoor human detection and outdoor vehicle to vehicle communication.}
    \label{AIGX-PHY6}
    \end{figure*}

   While there have been attempts to integrate GAI into wireless communication, they remain limited, especially when considering emerging technologies like extremely large-scale multiple-input-multiple-output (XL-MIMO), near-field communications, and integrated sensing and communication (ISAC)~\cite{cui2021integrating}. For instance, ISAC encompasses both communication and sensing modules, as shown in Fig.~\ref{AIGX-PHY6}, and each module has specific demands for bandwidth, power, and other resources. This complexity imposes new challenges in designing efficient wireless resource allocation strategies at the network layer to balance sensing and communication.
    
   Moreover, physical layer technologies such as antenna array and waveform design are also crucial for ISAC systems. For communication purposes, enhancing transmission reliability in multipath fading channels necessitates large antenna spacing to ensure independent signals across antennas. On the other hand, for sensing, estimating key parameters including the direction of arrival (DoA) of signal waves usually require antenna spacing to be less than or equal to half the wavelength to avoid ambiguities. These conflicting requirements introduce new challenges in the design of the antenna array for ISAC systems. Fortunately, the emerging of GAI and its recent applications in wireless communications, provide a promising way for resolving these dilemmas. Therefore, an in-depth investigation into the applications of GAI in ISAC systems, particularly in the physical layer, is imperative.

    Recognizing the challenges outlined above, this article conducts an extensive investigation on the application of GAI in the physical layer and the corresponding potential support for ISAC systems. Concretely, we first present an overview of five major GAI models and the ISAC system. After that, we thoroughly analyze the potential support of these GAI-enhanced physical layer technologies for ISAC from both sensing and communication perspectives. Finally, we provide a practical use case to explain how GAI can be used to tackle challenges in signal DoA estimation during sensing, a critical component of ISAC. Overall, the contributions of this article are summarized as follows.

\begin{itemize}
\item We conduct a review of five major GAI models and the ISAC system. Building on this, we analyze the potential applications of the GAI models in the physical, network, and application layers of the ISAC system, providing comprehensive insights of emerging sensing, localization, and communication technologies.

\item From different perspectives such as beamforming and signal detection, we investigate how GAI models enhance physical layer technologies. Subsequently, we analyze the support that these GAI-enhanced physical layer technologies provide for communication and sensing in ISAC systems, outlining technical issues and viable solutions. 

\item We propose a signal spectrum generator (SSG) to tackle the near-field DoA estimation problem when antenna spacing exceeds half the wavelength. Experimental results reveal that SSG yields a mean square error (MSE) of around 1.03 degrees in DoA estimation, confirming SSG's effectiveness while highlighting the importance of integrating GAI into the ISAC physical layer.

\end{itemize}

   \section{Overview of Generative AI and ISAC }
     This section first introduces the concepts of GAI and presents five representative GAI models. Following that, we introduce ISAC and generally explain GAI’s potential support for ISAC systems from the physical, network, and application layers.
     
    \subsection{Generative AI}
    GAI refers to a specific category of AI models that is trained on large datasets to learn the inherent patterns of the data distribution. Once trained, they can generate new data that is similar yet distinct from the training data, facilitating content production. Compared to the traditional AI models, the GAI models hold better ability to understand and capture the distribution of the complex and high-dimensional data~\cite{bond2021deep}. Hence, they find several applications across various fields. Among different GAI models, GANs, normalizing flows (NFs), variational autoencoders (VAEs), diffusion models (DFMs), and Transformers not only excel in generating digital content but also demonstrate significant applicability in the physical layer of wireless communications.

    \begin{itemize}
    \item GANs (Fig.~\ref{AIGX-PHY6}-part A) consist of a generator and a discriminator that compete during training, aiming for a particular equilibrium. The training is completed when the discriminator cannot differentiate between real and fake data. After that, the generator can produce similar, yet new data in a parallel manner. However, the training process is complex, as finding the equilibrium is harder than optimizing an objective function.
    
    \item NFs (Fig.~\ref{AIGX-PHY6}-part B) use invertible transformations to map basic distributions to target spaces for detailed analysis. These transformations create a flow that can be reversed, facilitating likelihood estimation. NFs can sample from complex probability distributions, which is useful for the unanalyzable data. However, many transformations may make the training process time-consuming.

    \item VAEs (Fig.~\ref{AIGX-PHY6}-part C) are neural networks designed to compress and reconstruct data. Unlike traditional autoencoders, VAEs can model the latent distribution and sample from the modeled space, benefiting data dimension reduction and feature extraction. Additionally, they can estimate the uncertainty in predictions and generate plausible outputs for a given input. However, generated samples are not always interpretable, as they are derived from the latent space.

    \item DFMs (Fig.~\ref{AIGX-PHY6}-part D) have attracted significant attention due to their image generation capabilities. During the training, DFMs corrupt training data with random noise and subsequently denoise the data to learn optimal hyperparameters. Once trained, they apply the learned parameters in the denoising process to generate samples. DFMs can be trained on incomplete datasets with a stable process, but inference requires many steps, making them less efficient for generating large datasets.

    \item Transformers (Fig.~\ref{AIGX-PHY6}-part E) are neural network architectures based on the self-attention mechanism, which can model long-range dependencies between elements in the input sequence and support parallel sequences processing, suitable for tasks involving substantial sequence data. Their design needs minimal inductive biases and is inherently suited for set-functions, enabling them to process multiple modalities using similar processing blocks.

    \end{itemize}

    Besides the above-mentioned models, there are some other GAI models, such as multimodal models, used in ChatGPT. These models possess strong data analysis and modeling capabilities, making them advantageous for incorporation into communication system designs.

    \subsection{Integrated Sensing and Communication}
    ISAC focuses on integrating wireless sensing and communication into a unified system. This aims at the efficient use of limited resources, while facilitating both functions~\cite{cui2021integrating}. From the physical layer, ISAC can be broadly classified into non-overlapping and overlapping systems. Specifically, non-overlapping systems include time-division, frequency-division, and space-division ISAC. For example, time-division ISAC allocates distinct signals to individual time slots for either sensing or communication tasks, allowing them to use their preferred waveforms. The overlapping systems can be divided into sensing-centric, communication-centric, and joint designs. For example, the communication-centric design can be achieved by appropriately modifying existing communication systems, and a representative example is Wi-Fi sensing. Compared to traditional wireless communication and sensing systems, the ISAC systems offer several advantages.
    \begin{itemize}
    \item Higher efficiency: By allowing communication and sensing to share resources, ISAC boosts the overall efficiency of wireless networks. 
    \item Lower cost: By eliminating the need for separate communication and sensing modules, ISAC lowers both hardware and power consumption costs for wireless devices.
    \item More versatile services: ISAC is capable of fulfilling users' communication requirements while concurrently offering sensing function, allowing it to deliver more services.
    \end{itemize}

    Benefiting from these advantages, ISAC systems can be applied across various scenarios and are thus considered one of the core technologies for future 6G networks.
    
    \subsection{Potential Applications of GAI in ISAC Systems}
    As aforementioned, we can see that GAI can serve ISAC systems from multiple perspectives. This can be broadly categorized into the physical, network, and application layers.

    \begin{itemize}
    \item \textbf{Physical layer:} GAI can be employed for channel estimation, anomaly signal identification, encoding, beamforming, etc, as shown in Fig.~\ref{AIGX-PHY6}. These GAI-enhanced physical layer technologies can improve the communication performance (e.g., reducing bit error rate (BER)) and enhancing the sensing accuracy (e.g., optimizing signal beams to increase target detection accuracy while avoiding interference in ISAC systems).

    \item \textbf{Network layer:} GAI can be utilized for designing resource allocation strategies, scheduling schemes, and incentive mechanisms, which could not only lower the system cost but also boost the operation efficiency. While methods such as deep reinforcement learning (DRL) are applicable here, GAI has been shown to be more effective in tasks like resource allocation~\cite{du2023beyond}.

    \item \textbf{Application layer:} GAI can be used to offer support in data generation, analysis, and feature extraction for various ISAC applications. This support not only facilitates in-depth analysis of communication or sensing data but also generates a substantial amount of data for both communication and sensing model training, which is difficult for other existing AI models.
    \end{itemize}

    In Table~\ref{GAIMD}, we summarize the above mentioned GAI models and their potential support for ISAC systems. Next, we detail GAI's applications in the physical layer.

\begin{table*}[ht]
\centering
\fontsize{6.8}{10}\selectfont    
 \caption{Five typical GAI models and corresponding potential support for ISAC at different layers.}
\begin{tabular}{>{\columncolor{blue!3}}p{2cm}| >{\columncolor{blue!2}}p{4cm}  >{\columncolor{blue!1}}p{5.5cm} p{2.6cm}}
    \hline
\diagbox [width=9em,trim=l] {GAI Models}{ Properties } &{\makecell[c]{Principles~\cite{bond2021deep}}}   &\makecell[c]{Advantages \& \textit{Disadvantages}}  &\makecell[c]{Potential applications~\cite{du2023beyond}}  \\

\hline
{\makecell[c]{GANs}} &{\makecell[l]{Train generators to produce  \\fake samples by competing \\with a discriminator}} &{\makecell[l]{a. Data generation in parallel manner\\
b. End-to-end training mechanism\\
\textit{c. Difficult to train as finding a Nash equilibrium} \\ \quad \textit{may be more difficult than optimizing the function} \\ \textit{d. Sensitive to hyperparameters}}}    &\multicolumn{1}{l}{\multirow{5}{*}{\makecell[l]{a. \textbf{Physical layer:} \\ \quad Channel estimation, signal detection and \\ \quad enhancement, joint source-channnel coding, \\ \quad beamforming, CSI compression, non-orthogonal \\ \quad multiple access (NOMA), secure transceiver \\ \quad design, synchronization, etc. \\ 
\\ b. \textbf{Network layer:} \\ \quad Resource (bandwidth, power, \\ \quad channel, etc) allocation strategy design,\\ \quad scheduling plan design (resource offloading, \\ \quad networking, crowdsourcing, etc), and incentive \\ \quad mechanism (auction, contract, etc) generation.  \\ \\ 
c. \textbf{Application layer:} \\ \quad ISAC data generation for model training, \\ \quad data repair and enhancement, high dimensional \\ \quad ISAC data distribution modeling and analysis, \\ \quad data feature extraction, data denoising and \\ \quad dimensionality reduction, etc. 
}}} \\

\hhline{|--- |}
 {\makecell[c]{NFs}} &{\makecell[l]{Use invertible ransformations to \\ convert simple distributions into \\ complex ones for data analysis and \\ generation}} &{\makecell[l]{a. Posterior distribution computing of latent variables\\ b. Understandable training process\\ c. \textit{Sensitive to the choice of the base distribution} \\ \textit{d. Struggle with the discrete and categorical data}}}   \\

\hhline{|--- |}
{\makecell[c]{VAEs}}  & {\makecell[l]{Compress and rebuild data by \\encoding it into a latent space\\ and then decoding it back to \\the original space}} &{\makecell[l]{a. Suitable for complex data processing (such as data \\ \quad  with high dimensional and complex distribution)
\\
b. Suitable for distributed training\\ \textit{d. Suffer from posterior collapse} \\ \textit{d. Require function to be continuous and differentiable}}} \\

\hhline{|--- |}
{\makecell[c]{DFMs}}  &{\makecell[l]{Learn optimal parameters by \\ adding noise to samples and generate \\ samples by applying these parameters \\ for denoising}} &{\makecell[l]{a. Flexible model structure (can be scaled to accommodate \\ \quad different levels of complexity)\\
b. Supports each step's probability distribution calculation \\ \textit{c. Low sampling rate (may need thousands of evaluation} \\ \quad  \textit{steps to draw a single sample)}}} \\ 

\hhline{|--- |}
{\makecell[c]{Transformers}}  &{\makecell[l]{Compute a weighted sum of the input \\ sequence elements, allowing more \\ attention on the important parts of the \\input sequence}} &{\makecell[l]{a. Capture long-range dependencies \\ 
b. Parallel processing of sequences \\ \textit{c. Difficult to interpret and visualize} \\ \textit{d. Limited ability to handle variable-sized inputs}}} \\ 
\hline
\end{tabular}\vspace{0cm}
    \label{GAIMD}
\end{table*}

\section{GAI-Enhanced Physical Layer Technologies for ISAC}
The physical layer includes several key technologies such as codebook design and channel estimation. In this section, we investigate how GAI strengthens various physical layer technologies and discuss their potential support for ISAC systems from both sensing and communication perspectives.

\subsection{From Communication Perspective}

\subsubsection{\textbf{Signal Detection}} Detecting signals in cases with unpredictable noise is challenging. NFs can infer latent variables, offering an effective solution. Hence, the authors in~\cite{he2021learning} propose a probabilistic machine-learning detection framework that employs NFs to approximate the unknown system noise in MIMO systems without any prior information. This approximation is driven by unsupervised learning with only noise samples, which is difficult to achieve with traditional AI models. Evaluations show that this framework not only stands out in terms of BER in environments with unanalyzable noise, but also reaches close to the maximum likelihood bound in environments with predictable noise. Besides NFs, other GAI models like GANs and VAEs can be also used for signal detection. In ISAC systems, the integration of communication and sensing creates more complex noise, additionally, differences in signal waveforms and other aspects between these two modules could exacerbate the issue. Therefore, NFs can also be employed to model the unknown noise, improving signal detection capability of ISAC systems.

\subsubsection{\textbf{Secure Transceiver Design}} The complexity of ISAC architectures and channel models complicates the design of security technologies. With the ability of processing complex data, VAEs can automatically manage codeword variation, which can be modeled as noise during transmission, making VAEs suitable for building secure transceiver pairs. In~\cite{lin2020variational}, the authors modify the VAE loss function at the receiver to include a security term, enhancing the receiver security. The unsupervised training is further used to strengthen the robustness against random codeword variations. In the case of imperfect CSI with the signal-to-noise ratio (SNR) range from -5 dB to 10 dB, the BER of this method at the eavesdropper is 0.05 higher than that of the autoencoder based on traditional neural networks. The same approach can be integrated into ISAC systems to enhance the security of the receiver and the robustness to codeword variations. However, when sensing and communication share the receiver, it is crucial to consider how adding the security term to a loss function might affect the sensing module.

\subsubsection{\textbf{Sparse Code Multiple Access}} In ISAC, various smart devices like unmanned aerial vehicles participate in communication and sensing, causing severe interference among devices. To mitigate this, combining GAI models with non-orthogonal multiple access (NOMA) techniques is a promising solution. The authors in~\cite{duan2023scma} introduce a GAN-based sparse code multiple access (SCMA) encoding and decoding approach. At the SCMA encoder, the generator is used to shorten the sequences in the information processing phase. Additionally, a noise layer is introduced to ensure a robust representation of the encoder output, thereby improving the noise immunity. At the decoder, PatchGAN serves as the discriminator to reduce both model parameters and computational load. Besides, an attention mechanism is inserted between the GAN's generator and discriminator to enhance the BER performance. Such designs can offer better connectivity of various smart devices involved in communication for ISAC, ensuring that control, scheduling, and other information can be timely transmitted to each device.

\subsubsection{\textbf{Joint Source-Channel Coding}} Coding is crucial for mitigating channel noise and interference, making it essential for communication of ISAC. Joint source-channel coding (JSCC) is an effective encoding method, but the complexity and discontinuity of the source data distribution present design challenges. To address this, in~\cite{saidutta2021joint}, the authors employ the VAE encoder to transform source data into a low-dimensional latent space and use the decoder to revert it to the original data for JSCC. During this process, one of multiple encoders is selected for transmission to tackle the issue of discontinuous projection. The evaluations show that the average peak SNR (PSNR) of the proposed method is nearly 3 dB higher than traditional methods based on convolutional neural networks. In ISAC systems where communication and sensing modules have independent encoding requirements and the channel is modeled as an additive Gaussian noise channel, such a method can directly contribute to the JSCC efficiency of communication module in ISAC.

\subsection{From Sensing Perspective}

\subsubsection{\textbf{CSI Compression}} Sensing in ISAC may need a significant amount of CSI from multiple antennas and frequencies, especially in systems like Wi-Fi based sensing. Hence, efficient compression, which facilitates the CSI storage and transmission, is essential. Given the superiority over traditional multi-layer perceptrons when output dimensionality far exceeds input, GANs are a preferred choice for CSI compression. In~\cite{tolba2020massive}, the authors use the CSiNet encoder at the transmitter to compress original CSI into a low-dimensional vector. Then, at the receiver, a deep convolutional GAN decoder reconstructs the original CSI from this compressed vector with the discriminator assessing its quality. The evaluations show that the normalized MSE of the proposed method is -7.05 dB, which is lower than -2.46 dB of CS-CsiNet based on deep learning, when the compression ratio is 1/64. Besides GANs, VAEs are also suitable for this task. These CSI compression methods show excellent reconstruction accuracy across varying compression ratios, providing support to reduce the overhead of CSI transmission and storage.

\subsubsection{\textbf{Beamforming}} Beamforming is a critical element in ISAC systems, significantly affecting the scanning accuracy in sensing tasks. Adaptive beam alignment remains a central challenge in this area. To address this, the authors in~\cite{hussain2021adaptive} introduce a VAE based dual timescale learning and adaptation framework. For the long timescale, a deep recurrent VAE (DR-VAE) is proposed to learn a probabilistic model of beam dynamics based on noisy beam-training observations. For the short timescale, an adaptive beam-training procedure is formulated as a partially observable Markov decision process. This is then optimized using point-based value iteration by leveraging both beam-training feedbacks and probabilistic predictions of the strongest beam pair provided by the DR-VAE. The proposed DR-VAE approach achieves near-optimal spectral efficiency, with a gain of 85\% over a conventional strategy that scans exhaustively over the dominant beam pairs. In ISAC, such a method not only minimizes the overhead associated with beam alignment during sensing process, but also boosts spectral efficiency, thereby increasing communication throughput.
\begin{table*}[]
\centering
\fontsize{6.5}{10}\selectfont 
\caption{The use of GAI in the physical layer and its potential support for communications of ISAC. Blue cells represent the discussed content, white cells reference other works, and empty cells denote unexplored areas.}
\begin{tabular}{p{1cm}|p{2cm}|p{2cm}|p{2cm}|p{2cm}|c}
\hline

\multirow{2}{*}{\diagbox [width=9em,trim=l] {Issues}{ Layers }}  & \multicolumn{4}{c|}{{Model layer}}  & \multicolumn{1}{c}{ISAC application layer}\\ \cline{2-6}                    

& \multicolumn{1}{c|}{GANs} & \multicolumn{1}{c|}{NFs} & \multicolumn{1}{c|}{VAEs} & \multicolumn{1}{c|}{DFMs}  & Communication \& sensing perspectives \\ \hline

\multicolumn{1}{c|}{Multiple access}  & \multicolumn{1}{l|}{\multirow{3}{*}{\cellcolor{blue!8}}{\makecell[l]{Source data enhancement\ \ \ }}}             &\multicolumn{1}{c|}{{\makecell[c]{--}}}             & \multicolumn{1}{c|}{--}               & \multicolumn{1}{c|}{--}   & \multirow{4}{*}{\makecell[l]{\textbf{Potential benefits for communication:} \\ a. Stronger signal detection capabilities in systems \\ \quad with unknown channel noise \\ b. More secure communication with lower BER \\ c. Better anomaly signal detection capabilities \\ d. Stronger spoofing signal generation and defense \\ \quad capabilities \\e. More efficient coding with higher PSNR \\f. Enhanced access capabilities for multiple devices}} \\  \hhline{|-----}

\multicolumn{1}{c|}{Signal detection} &\multicolumn{1}{l|}{{\makecell[l]{Learn the channel \\ transition probability}}} &\multicolumn{1}{l|}{\multirow{3}{*}{\cellcolor{blue!8}}{\makecell[l]{Model unanalyzable \\ system noise}}}   &\multicolumn{1}{l|}{{\makecell[l]{Learn the probability \\ distribution of the  \\ input signal}}}   &\multicolumn{1}{l|}{{\makecell[l]{Signal power spectral \\ density generation}}}  \\  \hhline{|-----}

\multicolumn{1}{c|}{\makecell[c]{Communication \\ security}} & \multicolumn{1}{l|}{Spoofing signal generation} & \multicolumn{1}{c|}{--} &\multicolumn{1}{l|}{\multirow{3}{*}{\cellcolor{blue!8}}{\makecell[l]{Handle the influence of \\ random codeword \\ variations}}}  & \multicolumn{1}{c|}{--} & \\  \hhline{|-----}

\multicolumn{1}{c|}{Coding} & \multicolumn{1}{l|}{Codebook design}  & \multicolumn{1}{c|}{--} &\multicolumn{1}{l|}{\multirow{3}{*}{\cellcolor{blue!8}}{\makecell[l]{Source data dimension \ \\ transformation}}}              &\multicolumn{1}{l|}{{\makecell[l]{Channel distribution \\ generation}}} & \\ \hline

\multicolumn{1}{c|}{CSI compression} & \multicolumn{1}{l|}{\multirow{3}{*}{\cellcolor{blue!8}}{\makecell[l]{CSI data compression and \ \  \\ decompression}}}  & \multicolumn{1}{c|}{--}  & \multicolumn{1}{c|}{--} & \multicolumn{1}{c|}{--}  & \multirow{4}{*}{\makecell[l]{\textbf{Potential benefits for sensing:} \\ a. Superior data compression ratio and improved \\ \quad reconstruction accuracy \\ b. Advanced CSI estimation accuracy for sensing \\ c. Enhanced beamforming performance with lower \\ \quad overhead for beam alignment \\ d. Repair and generate the sensing signal }} \\ \hhline{|-----}

\multicolumn{1}{c|}{Beamforming} &\multicolumn{1}{l|}{{\makecell[l]{Map the channels for \\ precoder extraction}}}  & \multicolumn{1}{c|}{--}  & \multicolumn{1}{l|}{\multirow{3}{*}{\cellcolor{blue!8}}{\makecell[l]{Learn the distribution \\ of the dynamic beams \ }}} & \multicolumn{1}{c|}{--} & \\ \hhline{|-----}

\multicolumn{1}{c|}{Channel estimation} &\multicolumn{1}{l|}{{\makecell[l]{Model the complex \\ channel distribution}}} & \multicolumn{1}{c|}{--}  &\multicolumn{1}{l|}{{\makecell[l]{Model unknown \\ channel distribution}}} & \multicolumn{1}{l|}{\multirow{3}{*}{\cellcolor{blue!8}}{\makecell[l]{Learn the distribution \\ of wireless channel}}} & \\ \hhline{|-----}

\multicolumn{1}{c|}{Signal enhancement} &\multicolumn{1}{l|}{{\makecell[l]{Synthetic micro-Doppler \\ spectrum signature}}} & \multicolumn{1}{c|}{--}  & \multicolumn{1}{c|}{--}   &\multicolumn{1}{l|}{{\makecell[l]{Produce and recover \\ the denoised channel}}} & \\ \hline
                        
\end{tabular}
\label{ISAC}
\end{table*}
\subsubsection{\textbf{Channel Estimation}} Channel estimation is important for sensing reliability, particularly in sensing systems that rely on CSI. Diffusion models, excel at learning high-dimensional gradients and model the log distribution of the data, are well-suited for modeling high-dimensional millimeter-wave MIMO channels. In~\cite{arvinte2022mimo}, the authors introduce a MIMO channel estimation method using score-based diffusion models. They first train a score-based generative model in an unsupervised manner using a database of known channels, which is independent of pilot symbols. Then, annealed Langevin dynamics is used for channel estimation by sampling from the posterior distribution. Compared to conventional supervised deep learning methods, this approach can offer a communication gain of up to 5 dB to the end-to-end coded communication system can reach up to 5 dB. More importantly, within ISAC systems, this approach holds the potential to solve the problems of estimating the channel in an out-of-distribution setting, i.e., the environments not seen during training, thereby providing more robust data support for the CSI-based sensing in complex channel conditions.

\subsubsection{\textbf{Signal Enhancement}} Signal parameter estimation is crucial for wireless sensing in ISAC systems, as it provides valuable observations for tasks like target detection and localization. Estimating signal parameters in low SNR conditions is particularly challenging. One effective strategy to address this issue is to improve the SNR using the generative capabilities of GAI models. Hence, in~\cite{cao2022pix2pix}, the authors convert low-SNR complex signals into images. Then, they employ a Unet structure as the GAN's generator to encode these images, effectively boosting the SNR. After that, PatchGAN, i.e., the discriminator, assesses the quality of the enhanced image. This approach successfully increases the SNR of the signal thereby yielding more accurate parameter estimation. Adapting this concept to ISAC, incomplete and low-SNR signals can be converted into images. GAI models, once trained, can then refine these images, effectively boosting the signal SNR and thereby improving parameter estimation and sensing performance. 

Besides the aforementioned applications, GAI can also be applied to sensing signal processing. For instance, in~\cite{lu2023radio}, the Transformer is used to capture inter-feature correlations among received signal strength observations, thereby boosting the multi-target localization capability. We summarize the above observations in Table~\ref{ISAC}.

\subsection{Discussion}

So far, various GAI models have been integrated into the physical layer, offering potential support for both the communication and sensing of ISAC systems from diverse perspectives. From the above investigations, we can see that the designs leverage the following prominent capabilities of GAI:

\begin{itemize}
\item \textbf{Capability of capturing complex data distributions.} For intricate datasets with complex distributions that are difficult, or even impossible to analyze directly, such as the noise and dynamic features of users, GAI models can be employed to capture their latent distributions. On this basis, the acquired distributions can be sampled, thereby supporting corresponding physical layer technologies, like signal detection in the system with complex noise and beam prediction in dynamic environments.

\item \textbf{Capability of transforming and processing data across various dimensions.} For high-dimensional data, GAI models can reduce its dimensionality through encoding and subsequently decode it to recover the original high-dimensional data. This facilitates the efficient compression, storage, and transmission of high-dimensional data within the ISAC system. For data with simpler distributions, GAI models can project them to more complex target spaces, thereby aiding in more efficient sampling and more accurate density estimation.

    \item \textbf{Capability of restoring and enhancing data}. For data in the ISAC system with a low SNR, such as the covariance matrix of received signals with low SNR as mentioned earlier, GAI models can effectively restore them. This restoration contributes to enhanced outcomes in subsequent stages, like more precise parameter estimation. Moreover, the generative capabilities of GAI can also recover incomplete data, ensuring that the subsequent processing can be effectively carried out.
    \end{itemize}

    \section{Case Study}
    Signal DoA estimation, which helps in identifying the location of the signal source, is crucial in both near-field and far-field ISAC systems. Besides, it also facilitates beamforming, enhancing the active near-field communication (NFC)~\cite{liu2023near}. However, when the antenna spacing exceeds half the wavelength (i.e., $\lambda$), the DoA estimation becomes challenging due to phase ambiguity. In this section, we show how to use GAI, i.e., diffusion models, to address this challenge, thereby providing support for near-field ISAC.
    
    \subsection{Problem Description}
\begin{figure*}[htp]
    \centering
    \includegraphics[width=1\linewidth]{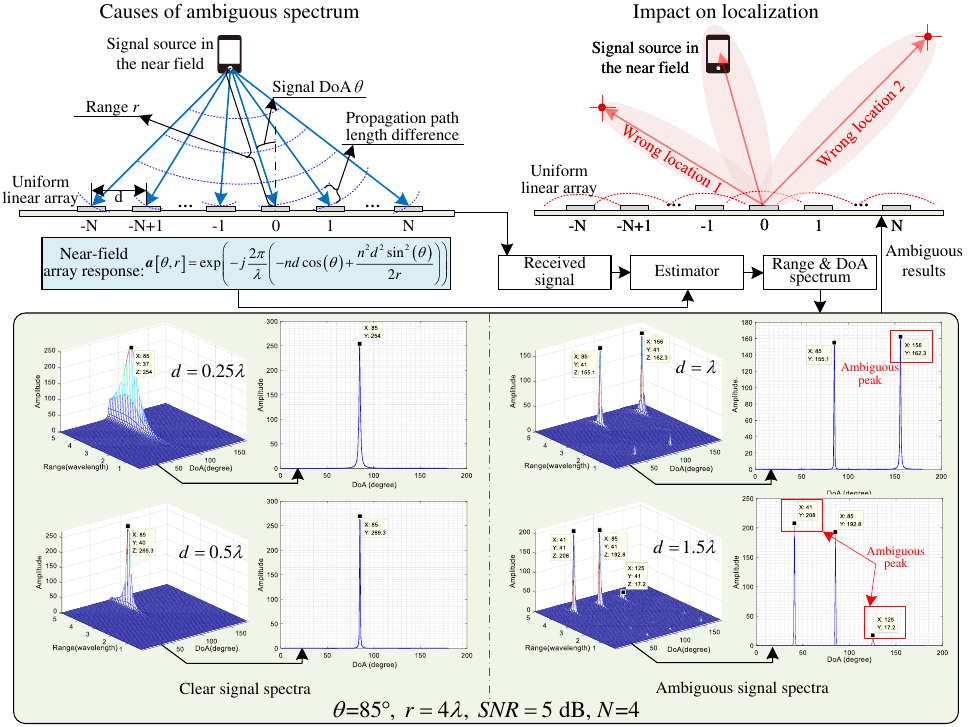}
    \caption{The cause of ambiguous signal spectrum and its impact on applications. Here, $\lambda$ is the signal wavelength, $d$ is the antenna spacing, $\theta$ is the DoA, $r$ is the distance between the signal source and the reference antenna, $2N+1$ is the total number of antennas. When $d$ is less than half of the $\lambda$, the signal DoA can be accurately estimated. However, as $d$ increases, for instance, to $\lambda$ or $1.5\lambda$, the signal spectrum becomes ambiguous, obstructing the identification of the true signal DoA and subsequently impacting further operations such as localization and beamforming."}
    \label{PBL}
    \end{figure*}

    Based on a uniform linear array (ULA), the DoA estimation relies on the phase difference across signals received by adjacent antennas in the array. This phase difference is a function of both the distance and direction of the signal source relative to the ULA. In the near-field context, the phase difference has a one-to-one correspondence with the DoA when the antenna spacing is less than or equal to $0.25\lambda$, allowing for effective DoA estimation, as shown by the clear signal spectrum in Fig.~\ref{PBL}. However, as the antenna spacing expands, for instance, to $\lambda$, the increased propagation path length causes the phase ambiguity, leading to an ambiguous spectrum, as shown in Fig.~\ref{PBL}. Under these conditions, the system is unable to identify the true signal DoA, which affects subsequent tasks like localization and beamforming.
    \subsection{Proposed Design}
\begin{figure*}[htp]
    \centering
    \includegraphics[width=1\linewidth]{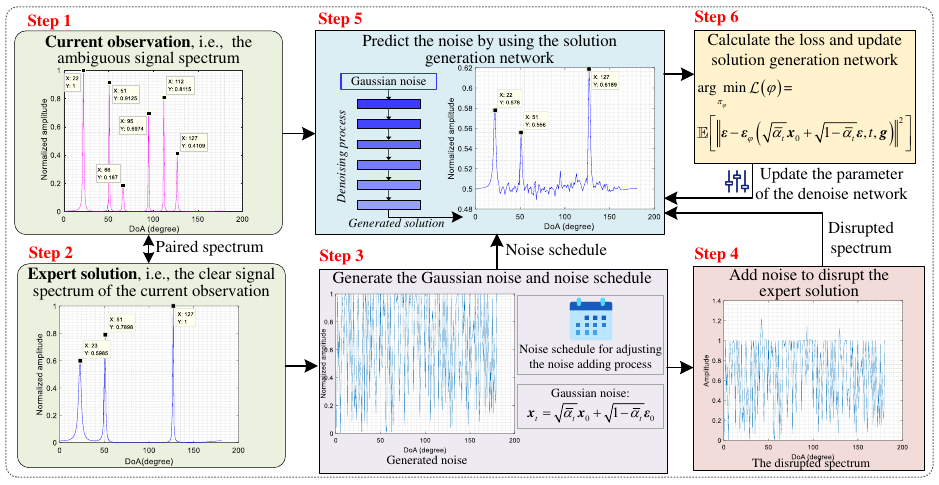}
    \caption{The structure of the proposed SSG. In Step 1, the current observation, i.e., the ambiguous signal spectrum, is obtained. In Step 2, the corresponding expert solution is obtained. Steps 3-6 detail the training process via forward and backward diffusion. Using the expert solution, the loss function is designed to minimize the discrepancy between the generated signal spectrum and the expert solution.}
    \label{SSG}
    \end{figure*}

    Essentially, the signal spectrum is a matrix and the distribution of data in it describes the signal DoA in the near-field. When the antenna spacing in the array exceeds half the wavelength, the signal spectrum becomes ambiguous, indicating a change in the data distribution and leading to an inability to recognize the correct signal DoA. The diffusion model's powerful inference capabilities can be employed to explore the relationship between ambiguous and clear signal spectra. Thereby, we propose a diffusion model-based signal spectrum generator (SSG), the architecture of which is illustrated in Fig.~\ref{SSG}.
    
    In the near-field scenario (shown in Fig.~\ref{PBL}) with $N=4$ and $d=0.5\lambda$, we produce 10,000 paired signal spectra via simulation, assigning 80\% for training and 20\% for testing. During the simulation, the number of signal sources is fixed at 3, with the corresponding DoAs and ranges randomly generated within 0-180 degrees and 0-6$\lambda$, respectively, and the SNR is also randomly generated between 0-5 dB. To ensure data consistency, the ambiguous spectra are obtained via DoA estimation using the signals captured by antennas with odd index (corresponding to an antenna spacing of $\lambda$), while signals from antennas 3 to 7 generate the corresponding correct spectra. Subsequently, the ambiguous spectrum serves as the observation, while the correct spectrum acts as the expert solution for training the SSG. During the training, the SSG adds noise to the expert solution and subsequently denoises it step by step, as shown in Steps 4 and 5, refining the denoising network hyperparameters along the way. These hyperparameters establish the denoising criteria, guiding the diffusion model's inference based on the observations. Therefore, after training, the SSG can generate the expert solutions (clear signal spectra) based on the given observations (ambiguous signal spectra) via the trained denoising network.

\subsection{Performance Evaluation}
\begin{figure*}[htp]
    \centering
   \includegraphics[width=1\linewidth]{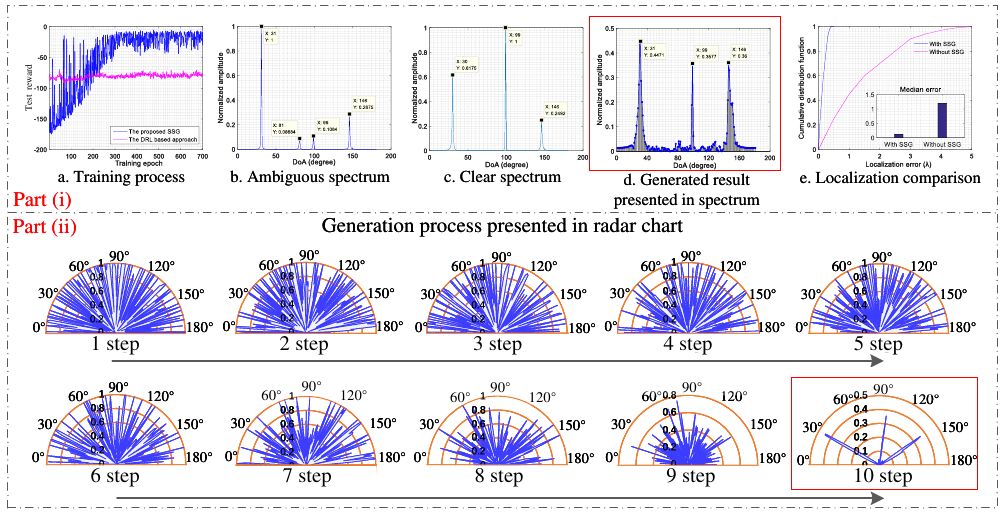}
    \caption{The experimental results. The Part (i) describes the training process of SSG as well as the comparison among the final generated signal spectrum, the observed ambiguous spectrum, and the real signal spectrum. The results presented in Part (ii) detail the signal spectrum generation process of the proposed SSG. During the inference process, the SSG starts with noise an uses the trained denoising network to denoise it. Therefore, as the number of inference steps increases, the noise in the spectrum gradually diminishes. Finally, after 10 inference steps, the clear signal spectrum is obtained.}
    \label{RSTS}
    \end{figure*}

The Part (i)-a in Fig.~\ref{RSTS} presents the test reward curve. According to the results, the difference between the solution generated by the SSG and the real expert solution gradually narrows over training, indicating that the SSG can learn the denoising network's hyperparameters through the noising and denoising processes and can subsequently utilize the denoising network to generate the corresponding expert solution. Furthermore, the test reward of SSG stabilizes around -10, better than -80 of DRL~\cite{du2023beyond} based method, indicating SSG is better than DRL in signal spectrum reconstruction. This may be due to the DRL struggling to focus on the key points in the spectrum corresponding to the DoAs, thereby failing to effectively learn the correct solution.

The results in Part (ii) of Fig.~\ref{RSTS} illustrate the expert solution generation process. As can be seen, the trained SSG can effectively produce the expert solution through sequential denoising based on the ambiguous spectrum in Fig.~\ref{RSTS} Part (i)-b. The generated signal spectrum in Part (i)-d depicts the DoAs of the three signal sources are 31, 99, and 146 degrees, which closely align with the expert solution, shown in Part (i)-c, of 30, 99, and 146 degrees, respectively. Based on the generated signal spectrum and the corresponding ground truth, we observe that the SSG achieves a DoA estimation MSE of about 1.03 degrees. This further proves that SSG can effectively produce the clear signal spectrum, which can be leveraged to both improve the energy efficiency of beamforming and reduce communication power consumption.

In addition, we investigate the impact of SSD on localization performance. During the test, we assume that the range is accurately estimated, and the system uses the DoA of the three peak points with the largest amplitude and the ranges to achieve localization. In Part (i)-e, results indicate a median localization error of about $1.25\lambda$ without using SSG. However, the use of SSG reduces this error to around $0.21\lambda$. This is intuitive, as ambiguous spectrums can lead the system to conduct localization using the incorrect DoA, causing notable errors.

  \section{Future Directions}
    \subsection{ GAI Application Security} 
    While GAI has demonstrated its potential in the physical layer, it also poses certain risks. For instance, attacks on the training datasets can lead to training non-convergence or even failure, thereby wasting significant computational resources. Attacks on GAI model itself could cause more severe consequences, such as ineffective channel estimation and coding, ultimately impacting the ISAC performance. Hence, future research should address these security issues from both the dataset and model perspectives. Blockchain technology can ensure data authenticity and provider reliability, while offering a unified management for multi-party data, hence serving as one effective approach to resolving these security issues.
    
    \subsection{Resource Allocation}
    The training and operation of GAI models consume computational, storage, and communication resources, disrupting the resource balance of the original system. Hence, integrating GAI models into the physical layer necessitates reallocating resources to ensure stable system operation. When local resources are abundant, strategies should be developed to maximize benefits while minimizing resource consumption based on task complexity and real-time requirements. When local resources are constrained, incentivization mechanisms, such as dynamic spectrum access, should be considered to ensure functional effectiveness, and then maximize benefits.

    \subsection{Cell Free ISAC}
    The decentralized architecture of cell-free massive MIMO effectively reduces the distance between the access point and the user, thereby minimizing path loss. This configuration is naturally conducive to the utilization of millimeter wave and terahertz frequencies for ISAC performance. Within this framework, GAI can be utilized to optimize factors such as precoding and combining. This integration has the potential to generate high-gain, narrow beams in a mobile cell-free setting, further enhancing the efficacy of both target tracking and high-capacity wireless fronthaul.


    \section{Conclusion}
     In this article, we investigated GAI's use in the physical layer from various perspectives. We concluded that these applications primarily leverage GAI's capabilities in complex data feature extraction, transformation, and enhancement. Subsequently, we analyzed how GAI-enhanced physical layer technologies can potentially support ISAC systems, considering both sensing and communication aspects. In the case study, we introduced the diffusion model based SSG. Operating in the physical layer, SSG addresses the DoA estimation problem that arises when array spacing exceeds half the wavelength. These insights emphasize the crucial role of GAI in the ISAC physical layer and the pressing need for a further exploration of its applications.
    
    \bibliographystyle{IEEEtran}
    \bibliography{IEEEabrv,Ref}
    \end{document}